\documentclass[11pt,aps,tightenlines,showpacs]{revtex4}
\usepackage{graphicx}
\usepackage{amssymb}
\usepackage{amsmath}

\begin{document}

\begin{center}
{\Large{\bf 
Muon capture on nuclei: 
random phase approximation \\
evaluation versus data for 6 $\le$ Z $\le$ 94 nuclei}}

\vspace{0.4cm}

{\large N. T. Zinner}

\vspace{0.2cm}

{\it Institute for Physics and Astronomy \\
University of Aarhus, DK-8000 Aarhus C, Denmark}

\vspace{0.2cm}
 
{\large K. Langanke} 

\vspace{0.2cm}

{\it GSI Darmstadt and Technische Universit\"at Darmstadt, 
D-64220 Darmstadt, Germany}

\vspace{0.2cm}

{\it and~} {\large P. Vogel}

\vspace{0.2cm}

{\it W. K. Kellogg Radiation Laboratory, 106-38 \\
California Institute of Technology, Pasadena, California 91125, USA}

\date{\today}

\vspace{0.3cm}

\begin{abstract}
\noindent We use the random phase approximation to systematically describe
the total  muon capture rates on all nuclei where they have been measured. 
We reproduce the
experimental values on these nuclei to better than 15\%
accuracy using the free nucleon weak form factors and  residual
interactions with a mild $A$ dependency. 
The isospin dependence and the effects associated with shell
closures are fairly well reproduced as well.
However, the calculated rates for the same residual
interactions would be significantly lower than the data if the in-medium
quenching of the axial-vector coupling constant is employed
to other than the true Gamow-Teller amplitudes. Our calculation thus
suggests that no quenching is needed 
in the description of semileptonic weak processes 
involving higher multipole transitions
and  momentum transfer $\sim m_{\mu}$, with obvious importance
to analogous weak processes.

\end{abstract}

\pacs{24.30.Cz, 23.40.-s, 23.40.Hc}
\end{center}

\maketitle

\section{Introduction}

\noindent The capture of a negative muon from the atomic $1s$ orbit,
\begin{equation}
\mu^- + (Z,N) \rightarrow \nu_{\mu} + (Z-1, N+1)^*
\end{equation}
is a semileptonic weak process that has been studied for a long time
(see, e.g., the recent review \cite{Mrev} or the earlier
one by Walecka \cite{Wal} and the classic by Mukhopadhyay
\cite{Muk} and the earlier references therein). The total capture
rate has been measured for many stable nuclei; in some cases the 
capture rates on separated isotopes have been
also determined \cite{Measday}.

The nuclear response in muon capture is governed by the momentum
transfer  of the order of the muon mass.  The phase
space and the nuclear response favor lower nuclear excitation
energies, thus the nuclear states in the giant resonance region
dominate. Since the experimental data are quite accurate, and the theoretical
techniques of evaluating the nuclear response in the relevant regime
are well developed, it is worthwhile to see to what extent the
capture rates are understood globally. Such a comparison may be viewed as
a general test of our ability to describe semileptonic weak
charged-current reactions with $q \sim m_{\mu}$ over a large range
of nuclei, where $q$ is the momentum transfer and $m_{\mu}=105.6$ MeV 
is the muon mass.

The present work represents a first fully comprehensive 
theoretical evaluation of the total muon capture rate over the full
range of nuclei where the experimental data are available. Previously,
the muon capture rates for selected nuclei encompassing, however,
a broad range of atomic charges, were calculated in Ref. \cite{Fearing}.
That paper was devoted mostly to the description of the radiative
muon capture, and the total muon capture rates were a byproduct with 
only a limited
agreement with the data. More along the lines of the present approach, Ref. \cite{Auer1} and 
\cite{Auer2} used a Hatree-Fock random phase approximation method and obtained good agreement 
with experiment, however only for a limited selection of nuclei.
The local Fermi gas model
was used successfully for the evaluation of the muon capture rate in selected
nuclei in Ref. \cite{Chiang}, and more recently in Ref. \cite{Nieves}. 

The present work is an extension of previous papers devoted to this issue
\cite{Kolbe1,Kolbe2}. In Ref. \cite{Kolbe1} the capture rates for $^{12}$C,
$^{16}$O and $^{40}$Ca were evaluated using the continuum random phase
approximation, and a very good agreement with the total rate was obtained. 
However, the residual interaction employed in \cite{Kolbe1} was adjusted
to describe other observables in the cases of $^{12}$C and
$^{16}$O. Moreover, it was necessary to quench the Gamow-Teller like (GT)
partial capture rate leading to the $1^+$ ground state of $^{12}$B.
In the later Ref. \cite{Kolbe2}
heavier nuclei with $N > Z$, $^{44,48}$Ca, $^{56}$Fe, $^{90}$Zr, 
and $^{208}$Pb
were also included with a similar success. 
In Ref. \cite{Kolbe2} it has also been shown that for the 
calculation of muon capture rates the
standard random phase approximation (SRPA) is essentially equivalent 
to the more computationally demanding continuum random phase 
approximation. Thus, the SRPA method is also
used in the present study. In Ref. \cite{Zinner04} the SRPA approach
has been used to study the muon capture rates to a long chain of calcium and
tin isotopes.

One of the important issues when evaluating the response of nuclei
to weak probes of relatively low energies is the problem of quenching
of the corresponding strength. The evidence for quenching comes primarily 
from the analysis of the beta decay of the $(sd)$ shell \cite{Wildenthal}
as well as $(p,f)$ shell \cite{pf} nuclei. In addition, the
interpretation of the forward angle $(p,n)$ and $(n,p)$ 
charge-exchange reactions
\cite{pn,Langanke95,Martinez96} leads to the same conclusion. All such evidence, so far,
is restricted to the GT strength and relatively low excitation energies. 
A convenient and customary way to account for this quenching is to use an
`effective' axial-vector coupling constant $g_A$, reducing it from its
nominal value of $g_A = 1.26$ to $g_A \sim 1$. 

The evaluation of the muon capture rate, reported here, suggests that
the quenching of $g_A$ is not needed to describe these data. (That conclusion
was already reached in Refs.\cite{Kolbe1,Kolbe2}.)
As stressed above, the mean excitation energy
in muon capture is in the region of giant resonances of
about 15 MeV (slowly decreasing with $A$ or $Z$),
and the GT-like operators contribute very little in heavier nuclei
where the neutrons and protons are in different oscillator shells.
In lighter nuclei, for $N$ and $Z$ less than 40, the GT strength
contributes, and is concentrated at  low energy. Thus, 
in agreement with the evidence mentioned above, we quench this, and
only this, part of the transition strength by a common factor 
$(0.8)^2 = 0.64$ \cite{Wildenthal}.
 
The present work, and the evaluation of the muon capture in general,
makes it possible to extend the study of quenching to higher multipoles,
and correspondingly to higher nuclear excitation energies. Such
processes, typically, depend primarily on the positions of the
corresponding giant resonances and on the overall strength. Our conclusions,
therefore, show that the SRPA method involving correlated
particle-hole excitations, is capable of describing
the inclusive semileptonic processes with momentum transfer $q$ 
of order $\sim m_{\mu}$ quite well. This is an important conclusion, applicable
to a variety of practically important subjects, e.g. detection of 
supernova neutrinos or evaluation of the nuclear matrix elements
for neutrinoless double beta decay.

The challenge of evaluating the muon capture rate in a wide variety 
of nuclei made it necessary to include several effects that were
not usually included in analogous calculations. Since we needed
to describe the bound muon in the $1s$ orbit well, we went beyond
the usual calculation of the muon density at the site of the
nucleus. First, we solved the Dirac equation in the field of the
finite size nucleus numerically. We then used its wave function, taking
into account that it is not constant between the origin and the
nuclear surface. For high $Z$ values the muon is relativistic,
and the `small' $p_{1/2}$ component of its wave function is nonnegligible.
As explained below, we used here (for the first time) the additional
transition matrix elements associated with that component.

Since our goal is to describe muon capture in all nuclei (except 
the very light ones) we have to desribe, at least crudely, effects
associated with the partial filling of the single-particle subshells
for nuclei that do not have magic numbers of protons and/or neutrons.
As described in the next section, we describe these effect by taking
into account the smearing of the proton and neutron Fermi levels
caused by pairing and deformation. It appears that this simplified 
treatment of complicated correlations, including those caused by
deformation, is sufficient for our purpose.

\section{Method and parameters}
\noindent In this calculation we have used the standard RPA model to describe
the nuclear excitations. In a previous work (\cite{Kolbe2}) this model
was shown to be just as good as the computationally more involved continuum
RPA. As residual interaction we use the phenomenological Landau-Migdal force.
For low mass nuclei the parameters for the force were taken from 
Ref. \cite{Speth1}. 
This choice was shown to be accurate in \cite{Kolbe1}.  
For muon capture
the most important term in the
Landau-Migdal force is the spin-isospin coupling constant g'. In \cite{Speth1}
the value  g'=0.7 is recommended, however, in a recent review \cite{Grummer1} 
g'=0.96 is used for heavy nuclei.
To accommodate this variation, we use an interpolation formula 
with a mild $A$ dependency:
\begin{equation}
g'=c_1+c_2\,A^{1/3}
\label{label:geq}
\end{equation}
where the constants $c_1$ and $c_2$ are fitted to yield g'=0.7 
in $^{16}$O and g'=0.96 in $^{208}$Pb.
We note that the change in the total capture rate in going 
from g'=0.7 to g'=0.96 is less than
10\%.

To get a basis of single-particle
states we diagonalize a Woods-Saxon potential (WSP) in a harmonic oscillator basis
of more than 8 major shells, thus enabling us to always have an excess of 
2$\hbar\omega$ of valence space above the Fermi level for both protons and neutrons.
As parameters of the WSP we use $R_0=1.2*A^{1/3}$ fm for the radius and 
$a=0.53$ fm for the diffuseness. The spin-orbit term is given as the derivative of the 
WSP times a strength $V_{so}$. Here we have simply used a fixed value of $V_{so}=-8.95$
through-out, initially checking that other choices did not significantly effect the 
total capture rate. To find the overall strength of the WSP, we fixed the last proton
and neutron particle energies to experimentally known masses. More specifically, for
a nucleus (A,Z) we found the energy of the last proton level from 
the proton separation energy $S_p$ in (A,Z).
For the last neutron level we used the neutron separation energy $S_n$,
but this time in the daughter (A,Z-1).

In order to be able to handle open-shell nuclei we have 
previously used a simple scheme 
where partial occupancies was treated by multiplying the open 
level matrix elements by
occupation numbers corresponding to an independent particle model
~\cite{Kolbe3}. In this work we have attempted to improve on 
this
treatment by solving the standard BCS equations to determine 
the occupation numbers.
Following Ref.\cite{Fetter1} the occupation probabilities are given by
\begin{equation}
v_{k}^{2}=\frac{1}{2}\left( 1- \frac{\epsilon_k-\mu}{
\sqrt{\Delta^2+(\epsilon_k-\mu)^2}}\right)
\end{equation}
where $\epsilon_k$ are the single-particle energies and the chemical potential $\mu$ is
fixed by the condition $N=\sum_k v_{k}^{2} (2j_k + 1)$. 
The pairing gap $\Delta$ is obtained by
the procedure descibed in Ref. \cite{Bohr1}.

The formalism used to evaluate the total muon capture rate is 
that of Ref. \cite{Wal}. As mentioned 
in the Introduction, we treat the muon wavefunction by solving the Dirac equation
in the extended charge of the nucleus, which is assumed to be of the Woods-Saxon form with 
the same paramteres as above. For nuclei with large values of $Z$, the atomic binding energy
becomes a significant fraction of the muon rest mass and the small component of the
Dirac bi-spinor may not be negligble in this range. We therefore explicitly include all terms
containing both large and small components in our transition operators. An outline of the complications
arising from this is given in the Appendix.

\section{Results}

\begin{figure}[ht!]
\centering
\includegraphics[width=0.95\textwidth,clip=true]{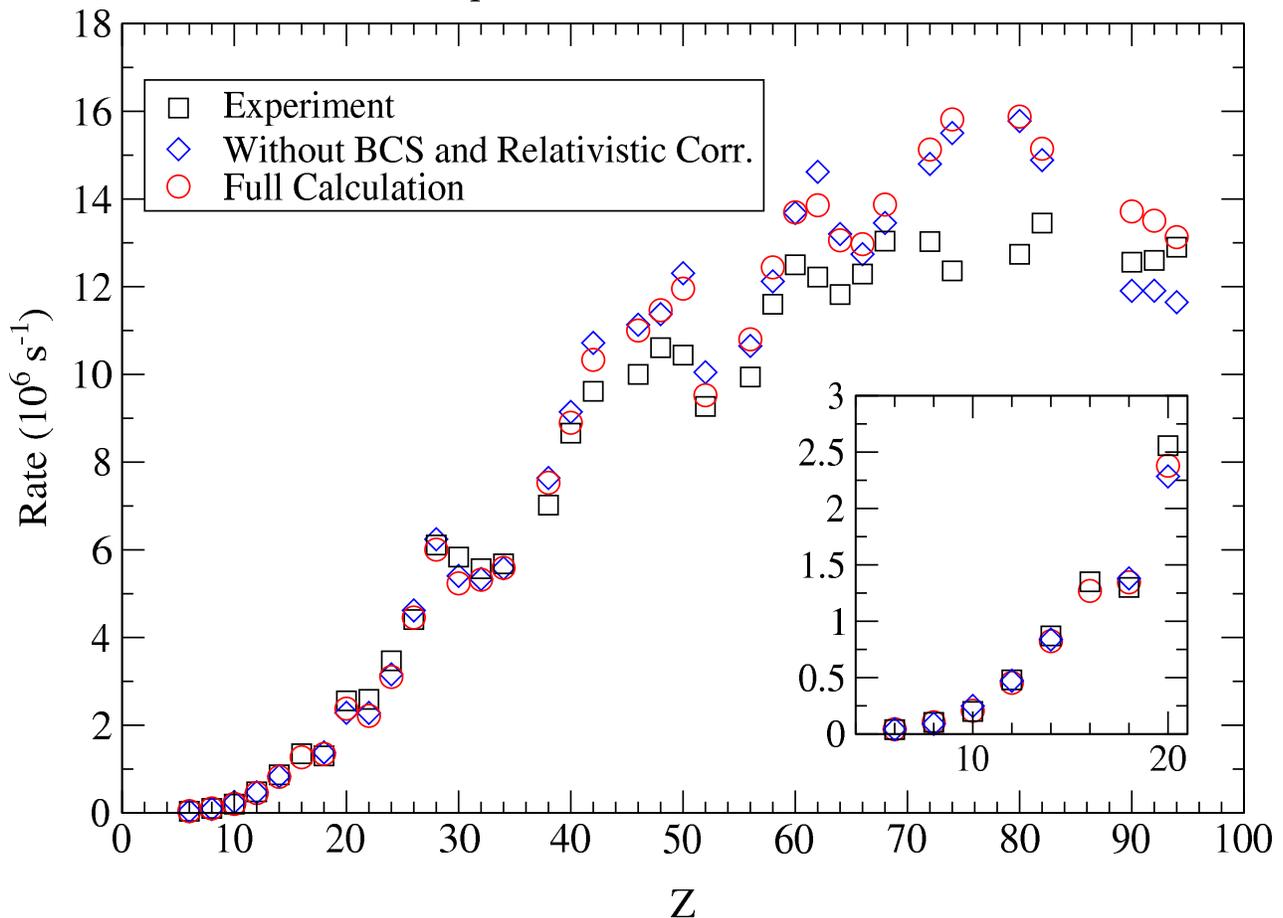}
\caption{(color online) Comparison between the measured total muon capture rates
\cite{Measday} denoted by squares $\Box$, the calculated rates with all
corrections (empty circles $\bigcirc$), 
and the calculated rates without the BCS and relativistic corrections 
(diamonds $\Diamond$ ). 
The insert, in larger scale, shows the same results for light
nuclei. When the measurements
are for the natural abundance of a given element, the calculation 
represent the corresponding combination of the individual isotopes.}
\label{fig:final}
\end{figure}

\begin{figure}[h!]
\centering
\includegraphics[width=0.95\textwidth,clip=true]{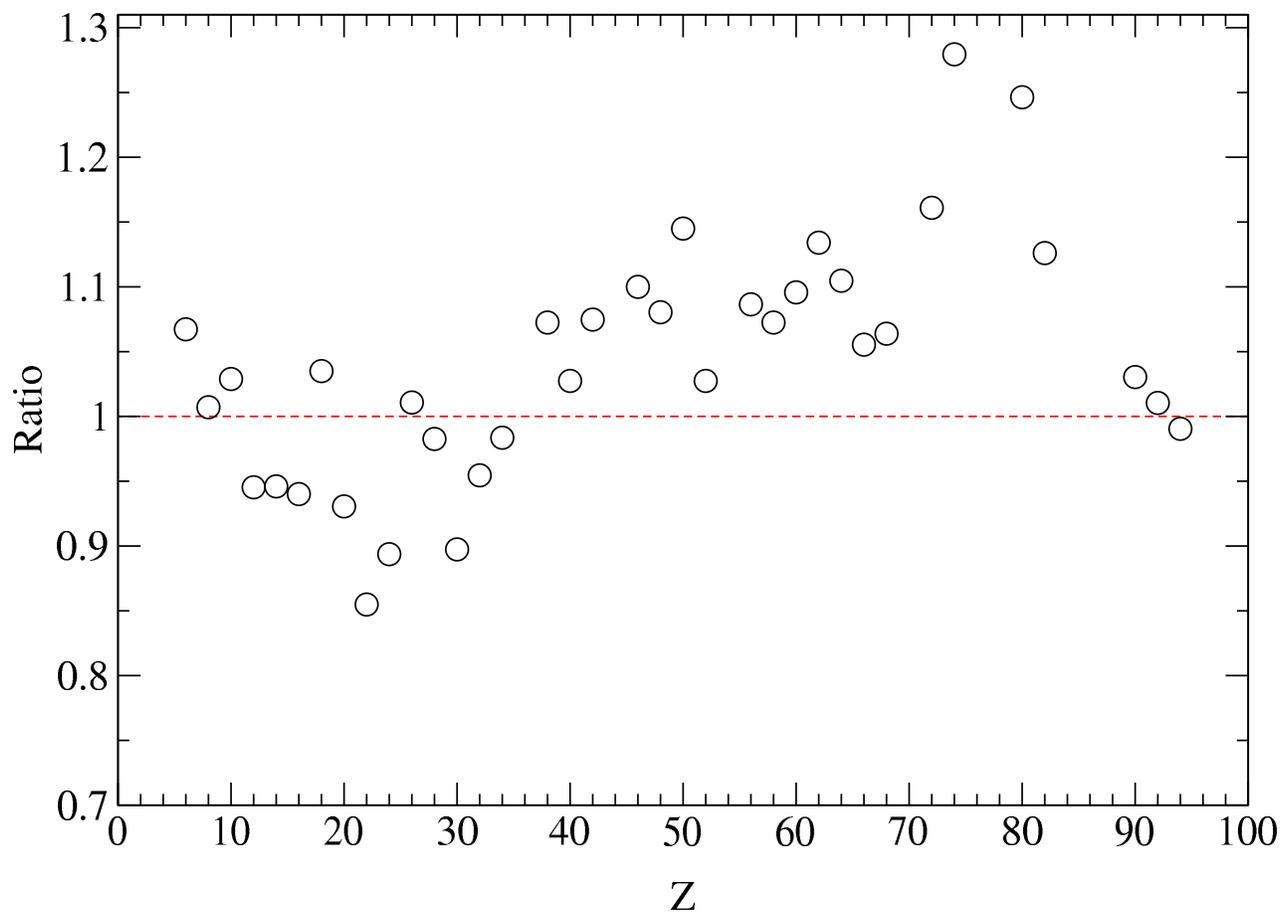} 
\caption{(color online) Ratios of the calculated and measured total muon capture
rates vs. the atomic number.}
\label{fig:ratio}
\end{figure}

\noindent Our calculated total muon capture rates for all nuclei for which measured
values exist are shown in Figure~\ref{fig:final}. One sees that the overall agreement is quite good. 
With the exceptions at  $Z=74$ and $80$ the calculations reproduce the experimental values
to 15\% or better. In Figure~\ref{fig:ratio} we 
provide a ratio plot where the degree of agreement is better seen. 
In addition, in Table \ref{tab:rates} we collect all our calculated 
capture rates,
including the results for individual isotopes.

\begin{table}[h!]
\begin{center}
\begin{tabular}{|l|c|c||c|c|c||c|c|c|}
\hline
Nuc & exp & calc & Nuc & exp & calc & Nuc & exp & calc \\
\hline
$^{12}$C  & 0.039            & 0.042  & 
$^{16}$O  & 0.103            & 0.104  &
$^{18}$O  & 0.088            & 0.089 \\
$^{20}$Ne & 0.204            & 0.206 &
$^{24}$Mg & 0.484            & 0.454  & 
$^{28}$Si & 0.871            & 0.823  \\
$^{32}$S  & 1.352            & 1.269  & 
$^{40}$Ar & 1.355            & 1.345  &
$^{40}$Ca & 2.557            & 2.379  \\ 
$^{44}$Ca & 1.793            & 1.946  &
$^{48}$Ca & 1.214$^1$        & 1.455  & 
$^{48}$Ti & 2.590            & 2.214  \\
$^{nat}$Cr& 3.472            & 3.101  & 
$^{50}$Cr & 3.825            & 3.451  &
$^{52}$Cr & 3.452            & 3.085  \\ 
$^{54}$Cr & 3.057            & 3.024  &
$^{56}$Fe & 4.411            & 4.457  & 
$^{nat}$Ni& 5.932            & 6.004  \\
$^{58}$Ni & 6.110            & 6.230  & 
$^{60}$Ni & 5.560            & 5.563  &
$^{62}$Ni & 4.720            & 4.939  \\ 
$^{nat}$Zn& 5.834            & 5.235  &
$^{64}$Zn &                  & 5.735  & 
$^{66}$Zn &                  & 4.976  \\
$^{68}$Zn &                  & 4.328  & 
$^{nat}$Ge& 5.569            & 5.317  &
$^{70}$Ge &                  & 5.948  \\ 
$^{72}$Ge &                  & 5.311  &
$^{74}$Ge &                  & 4.970  & 
$^{nat}$Se& 5.681            & 5.588  \\
$^{78}$Se &                  & 6.023  & 
$^{80}$Se &                  & 5.485  &
$^{82}$Se &                  & 5.024  \\ 
$^{nat}$Sr& 7.020            & 7.529  &
$^{86}$Sr &                  & 8.225  & 
$^{88}$Sr & 6.610            & 7.445  \\
$^{nat}$Zr& 8.660            & 8.897  & 
$^{90}$Zr &                  & 8.974  &
$^{92}$Zr &                  & 9.254  \\ 
$^{94}$Zr &                  & 8.317  &
$^{nat}$Mo& 9.614            & 10.33  & 
$^{92}$Mo &                  & 10.80  \\
$^{94}$Mo &                  & 11.01  & 
$^{96}$Mo &                  & 10.04  &
$^{98}$Mo &                  & 9.153  \\ 
$^{nat}$Pd& 10.00            & 11.00  &
$^{104}$Pd&                  & 12.71  & 
$^{106}$Pd&                  & 11.44  \\
$^{108}$Pd&                  & 10.44  & 
$^{110}$Pd&                  & 9.607  &
$^{nat}$Cd& 10.61            & 11.46  \\ 
$^{110}$Cd&                  & 12.58  &
$^{112}$Cd&                  & 11.51  & 
$^{114}$Cd&                  & 11.21  \\
$^{116}$Cd&                  & 10.44  & 
$^{nat}$Sn& 10.44            & 11.95  &
$^{116}$Sn&                  & 13.08  \\ 
$^{118}$Sn&                  & 12.35  &
$^{120}$Sn&                  & 11.64  & 
$^{122}$Sn&                  & 10.82  \\
$^{124}$Sn&                  & 10.15  & 
$^{nat}$Te& 9.270            & 9.523  &
$^{126}$Te&                  & 10.20  \\ 
$^{128}$Te&                  & 9.639  &
$^{130}$Te&                  & 9.043  & 
$^{nat}$Ba& 9.940            & 10.80  \\
$^{136}$Ba&                  & 11.45  & 
$^{138}$Ba&                  & 10.73  &
$^{nat}$Ce& 11.60            & 12.44  \\ 
$^{140}$Ce&                  & 12.38  &
$^{142}$Ce&                  & 12.95  & 
$^{nat}$Nd& 12.50            & 13.70  \\
$^{142}$Nd&                  & 13.67  & 
$^{144}$Nd&                  & 14.12  &
$^{146}$Nd&                  & 13.15  \\ 
$^{nat}$Sm& 12.22            & 13.86  &
$^{148}$Sm&                  & 15.01  & 
$^{152}$Sm&                  & 13.23  \\
$^{154}$Sm&                  & 12.08  & 
$^{nat}$Gd& 11.82            & 13.06  &
$^{156}$Gd&                  & 14.15  \\ 
$^{158}$Gd&                  & 13.06  &
$^{160}$Gd&                  & 12.03  & 
$^{nat}$Dy& 12.29            & 12.97  \\
$^{162}$Dy&                  & 13.45  & 
$^{164}$Dy&                  & 12.54  &
$^{nat}$Er& 13.04            & 13.87  \\ 
$^{166}$Er&                  & 14.46  &
$^{168}$Er&                  & 13.51  & 
$^{170}$Er&                  & 13.22  \\
$^{nat}$Hf& 13.03            & 15.13  & 
$^{178}$Hf&                  & 15.44  &
$^{180}$Hf&                  & 14.89  \\ 
$^{nat}$W & 12.36            & 15.81  &
$^{182}$W &                  & 16.37  & 
$^{184}$W &                  & 15.79  \\
$^{186}$W &                  & 15.32  & 
$^{nat}$Hg& 12.74            & 15.88  &
$^{198}$Hg&                  & 17.17  \\ 
$^{200}$Hg&                  & 16.29  &
$^{202}$Hg&                  & 15.43  & 
$^{204}$Hg&                  & 14.58  \\
$^{nat}$Pb& 13.45            & 15.15  & 
$^{206}$Pb&                  & 15.54  &
$^{208}$Pb&                  & 14.97  \\ 
$^{232}$Th& 12.56            & 13.71  &
$^{234}$U & 13.79            & 14.89  & 
$^{236}$U & 13.09$^2$        & 14.17  \\
$^{238}$U & 12.57$^2$        & 13.51  &
$^{242}$Pu& 12.90            & 13.13  & 
$^{244}$Pu& 12.40$^3$        & 12.70  \\
\hline
\end{tabular}
\end{center}

\caption{Calculated rates for natural elements and for the important individual isotopes. 
All rates are in the form $xx \times 10^{6}$ s$^{-1}$.
If the natural abundance has more than 90\% of a given isotope then the Nuc column gives this particular 
isotope and the calculation is done for this nucleus only. Whenever a nucleus has the superscript $^{nat}$ 
the calculation is a weighted combination of all isotopes contributing more than 10\% to the natural 
abundance.
An empty experiment box means that the given isotope has not been measured. Natural 
abundance measurements appear in the box next to the most abundant isotope.
Experimental data are from Ref. \cite{Measday}, where the original sources may be found.
$^1$ is taken from Ref. \cite{ca48}, $^2$ is from Ref. \cite{Hanscheid} 
and $^3$ is from Ref. \cite{David}}
\label{tab:rates}
\end{table}

In Fig. \ref{fig:final} the total capture rates are also compared with values 
obtained from calculations where 
the BCS occupancy and the relativistic corrections are turned off. 
From this comparison it is clear that these are only small 
corrections, providing justification for the calculation done 
in Ref. \cite{Kolbe1,Kolbe2,Kolbe3}.
At low $Z$ one especially 
notices the good reproduction of the distinct dips in the rates above the magic 
numbers $Z=20$ and $28$. For $Z=50$ 
the same trends is visible in both calculation and experiment, but the calculations  overshoot the 
experimental values somewhat. Just above $Z=60$, where the $N=82$ 
closed neutron shell comes into play, we also overestimate 
the capture rates. This continues into the region with $Z=74$ and $80$, 
and also, to a lesser extent, to the doubly-magic nucleus 
$^{208}$Pb.  One should  remember that 
some of the nuclei above $Z=50$ and below $Z=82$ are 
deformed and thus have  different 
single-particle structures than the ones given by our 
spherical mean-field model, thus a perfect
agreement should not be expected. The fact that the calculated values are again 
approaching experiment at Th,U and Pu is likely a consequence of the same dip after the magic shell closure that was seen at lower $Z$ 
also (these trends were also noted in Ref. \cite{Hanscheid}, see in particular figure 3 of that 
reference).

As stated above, we use the unquenched value for the axial-vector coupling 
constant for all multipole operators, except for the true Gamow-Teller
transition. For most of the light- and medium-mass nuclei,
$\lambda=1^-$ (dipole-like) and $\lambda=2^+$ (quadrupole-like) transitions
dominate. However, for $^{208}$Pb and the heavier nuclei, $\lambda=1^+$
transitions contribute significantly to the total capture rate.
For these nuclei, the neutron excess is already so large that,
in the simple independent particle model, 2 major oscillator shells
have to be overcome when changing a proton into a neutron
in the muon capture process. Thus, for these excitations the $\lambda=1^+$
multipole transition corresponds to a 2$\hbar \omega$ mode and not to
a (0 $\hbar \omega$) Gamow-Teller transition; the contribution
of the latter to the rate in the heavy nuclei vanishes in our calculations. 
We have not renormalized the axial-vector coupling constant
for such 2 $\hbar\omega$ $1^+$ transitions, supported by the good
agreement of the
the calculated capture rates with the measured results for
the heavy nuclei.

An important test of the ability of a nuclear structure model to reproduce the data
is the dependence of the muon capture rate on the number of neutrons for a fixed
nuclear charge $Z$. Typically, the rate decreases with increasing $N$ (or $A$)
as subsequently more neutron levels are getting blocked. This effect
is incorporated into the well-known Primakoff parametrization by its (N-Z)
dependence \cite{Primakoff}. As examples,
table \ref{tab:rates} include  three isotope chains of nuclei (Ca, Cr, and
Ni),
where total capture rates for
individual isotopes have been measured. One sees that the isotope
dependence is well reproduced by
our calculations. Analogous calculations were performed
in Ref~. \cite{Kolbe4,Zinner04}.

\section{Conclusion}

\noindent The present analysis shows that the standard random phase approximation 
method is capable of describing quite well the total $\mu^-$ capture
rates for essentially all stable nuclei. The dependence of the capture
rate on the isospin, or neutron excess, the so-called Primakoff rule
\cite{Primakoff}, is also fairly well, albeit not perfectly, 
reproduced. Our calculation even
describes the rather subtle effects of shell closures when considering
the dependence of the capture rate on $Z$ and/or $A$. There is no 
indication of the need to apply any quenching to the operators responsible
for the muon capture, in particular those involving 
single-particle transitions from one oscillator shell to another, 
i.e. other than those involving $0 \hbar \omega$ spin and isospin
changing operators.

Given the task of describing the capture in a variety of nuclei, including
those with high charge $Z$ and nuclei with unfilled shells, it became
necessary to consider several effects that have not been typically
included previously. One of them is a fully relativistic treatment
of the muon bound state, including the effects associated with
the `small' $p_{1/2}$ component of its wave function. Another one
is the effect of the smearing of the Fermi level (both for
protons and neutrons) in nuclei that have nonmagic $Z$ or $N$ numbers. 
Even though the corresponding corrections are not
very large, they contribute noticeably to the overall
good agreement between the experimental data and our calculated values. 

Our findings then can be used as guidance in the evaluation of a wide
variety of semileptonic weak processes on nuclei with  similar
momentum transfer.  

\section{Acknowledgment}

\noindent One of us (P.V.) gratefully acknowledges the hospitality 
at GSI, where part of this work was done. 

\newpage
\section{Appendix}
\noindent In this appendix we give an outline of  terms arising from the inclusion of the 
fully relativistic treatment of the bound muon wave function, i.e., of   both the large and 
small components. The muon 
wavefunction we use has the form
\begin{equation}
\phi_{jm}=\left(\begin{array}{ccccc} i g(r)\Omega_{jlm}(\hat{r}) \\
 -f(r) \Omega_{jl'm}(\hat{r}) \end{array}  \right)
\end{equation}
where the radial functions satisfy the equations
\begin{equation}
\frac{\textrm{d}}{\textrm{d}r}
\left( \begin{array}{cc} g \\ f \end{array} \right) =
\left( \begin{array}{cccccc} -\frac{1+\kappa}{r} & E+m_\mu-V(r) \\ -E+m_\mu+V(r) & -
\frac{1-\kappa}{r}\end{array} \right)
\left(\begin{array}{cc} g\\ f\end{array}\right),
\end{equation}
where $g(r),f(r)$ are the large and small components, respectively. Here 
\begin{equation}
\kappa=\left\{\begin{array}{lll} -(l+1) & ,~ j=l+\frac{1}{2} \\ l &
,~ j=l-\frac{1}{2}\end{array}\right\}
\end{equation}
and
\begin{equation}
\Omega_{j,l,m}(\hat{r})=\sum_{m_l,m_s}(l,1/2,j\vert
m_l,m_s,m)Y_{l,m_l}(\hat{r})\chi_{m_s}
\end{equation}
These equations are entirely general and can be found in e.g. Ref. 
\cite{greiner1}. Since we assume that the muon is captured from the 
atomic $1s$ orbit we have $l=0$ and $l'=1$. Since $l'$ is non-zero, 
we can no longer just multiply the wavefunction with the 
irreducible nuclear operators and obtain good 
total angular momentum. If the positive z-axis 
is chosen along the direction of the out-going neutrino then 
its wavefunction becomes
\begin{eqnarray*}
\psi = N \left( \begin{array}{ccc} \chi_- \\ -\chi_-\end{array}  \right)
\exp(i \vec{p}\cdot\vec{x})
\end{eqnarray*}
Here $\chi_-$ are the usual spin down Pauli two-spinors and N is a
normalization given in \cite{greiner1}. This wavefunction makes the neutrino purely left-handed 
as the standard model prescribes.

The approach used in \cite{Wal} neglects the small component in the muon wave
function 
and expands the neutrino plane wave in 
multipoles. As the large component has $l=0$, angular momentum coupling 
of the muon wave function and the multipole operators is quite straightforward.
This is, however, no longer the case if the small component with
orbital angular momentum $l=1$ is considered, implying the need
for a cumbersome recoupling of angular momenta to
regain tensor operators that can be applied in the nuclear Hilbert space. 
This results in a more complicated 
expression for the weak Hamiltonian governing  muon capture with 
several new terms. 
In the notation used in \cite{Wal} the Hamiltonian with all terms 
from both components can be written as 
\begin{eqnarray*}
&\phantom{e}&\mathcal{H}=\\
&\phantom{e}& =\frac{2\,{G_F\, \cos \theta_C\, N^*}}{\sqrt{2}}[\sum_{J=0}^{\infty}\sqrt{4\pi}[J](-i)^J\{ 
i \delta_{m,-1/2}
\left\{\mathcal{M}_{J,0}^{'}-\mathcal{L}_{J,0}^{'}\right\}\\
&\phantom{\int}& +\alpha\left(J-1,J,m+\tfrac{1}{2}\right)\mathcal{T}_1\left(J-1,J,m+\tfrac{1}{2}\right)\\
&\phantom{\int}& +\alpha\left(J+1,J,m+\tfrac{1}{2}\right)\mathcal{T}_1\left(J+1,J,m+\tfrac{1}{2}\right)\\
&\phantom{=}&-i\beta_+(J,J,m)\mathcal{T}_2(J,J,m+\tfrac{1}{2})\\
&\phantom{=}&-i\beta_+(J+1,J,m)\mathcal{T}_2(J+1,J,m+\tfrac{1}{2})\\
&\phantom{=}&-i\beta_+(J+1,J+2,m)\mathcal{T}_2(J+1,J+2,m+\tfrac{1}{2})\\
&\phantom{=}&-i\beta_-(J-1,J-2,m)\mathcal{T}_3(J-1,J-2,m+\tfrac{1}{2})\\
&\phantom{=}&-i\beta_-(J-1,J,m)\mathcal{T}_3(J-1,J,m+\tfrac{1}{2})\\
&\phantom{=}&-i\beta_-(J,J,m)\mathcal{T}_3(J,J,m+\tfrac{1}{2}) \}\\
&\phantom{e}&+\sum_{J=1}^{\infty} \sqrt{4\pi}[J](-i)^J \{i \delta_{m,1/2}
\left\{\mathcal{J}^{'el}_{J,1}-\mathcal{J}^{'mag}_{J,1}\right\}\\
&\phantom{e}& -\delta(J-1,J-1,m)\mathcal{T}_4(J-1,J-1,m+\tfrac{1}{2})\\
&\phantom{e}& -\delta(J-1,J-1,m)\mathcal{T}_4(J-1,J-1,m+\tfrac{1}{2})\\
&\phantom{e}& -\delta(J,J,m)\mathcal{T}_4(J,J,m+\tfrac{1}{2})\\
&\phantom{e}& -\delta(J,J+1,m)\mathcal{T}_4(J,J+1,m+\tfrac{1}{2})\\
&\phantom{e}& -\delta(J+1,J+1,m)\mathcal{T}_4(J+1,J+1,m+\tfrac{1}{2})\\
&\phantom{e}& +i\eta_+(J,J,m)\mathcal{T}_2(J,J,m+\tfrac{1}{2})\\
&\phantom{e}& +i\eta_+(J+1,J,m)\mathcal{T}_2(J+1,J,m+\tfrac{1}{2})\\
&\phantom{e}& +i\eta_+(J+1,J+2,m)\mathcal{T}_2(J+1,J+2,m+\tfrac{1}{2})\\
&\phantom{e}& -i\eta_-(J-1,J,m)\mathcal{T}_3(J-1,J,m+\tfrac{1}{2})\\
&\phantom{e}& -i\eta_-(J,J,m)\mathcal{T}_3(J,J,m+\tfrac{1}{2})\\
&\phantom{e}& -i\eta_-(J-1,J-2,m)\mathcal{T}_3(J-1,J-2,m+\tfrac{1}{2})\}]
\end{eqnarray*}
Here we have defined the tensor operators in the nuclear Hilbert space as
\begin{eqnarray*}
\mathcal{M}_{J,M}^{'}&=&\int \textrm{d}^{3} \vec{x} g(r)Y_{0,0} j_J(\kappa x) Y_{J,M} J_0\\
\mathcal{L}_{J,M}^{'}&=&\frac{i}{\kappa}\int \textrm{d}^{3} \vec{x} g(r)Y_{0,0}
\nabla\left(j_J(\kappa x) Y_{J,M}\right) J_0\\
\mathcal{J}^{'mag}_{J,M}&=&\int \textrm{d}^{3}
\vec{x} g(r)Y_{0,0}j_J(\kappa x)\vec{\mathcal{Y}}^{M}_{J,J,1}\cdot\vec{J}\\
\mathcal{J}^{'el}_{J,M}&=&\frac{1}{\kappa}\int
\textrm{d}^{3} \vec{x} g(r)Y_{0,0}\nabla\wedge\left(j_J(\kappa x)\vec{\mathcal{Y}}^{M}_{J,J,1}\right)\cdot\vec{J}\\
\mathcal{T}_1(\gamma,\rho,\mu)&=&\int \textrm{d}^{3} \vec{x} f(r) j_{\rho}(\kappa x) Y_{\gamma,\mu}J_0\\
\mathcal{T}_2(\gamma,\rho,J,\mu)&=&\int \textrm{d}^{3} \vec{x} f(r) j_{J+1}(\kappa x)
\vec{\mathcal{Y}}^{\mu}_{\gamma,\rho,1}\cdot\vec{J}\\
\mathcal{T}_3(\gamma,\rho,J,\mu)&=&\int \textrm{d}^{3} \vec{x} f(r) j_{J-1}(\kappa x)
\vec{\mathcal{Y}}^{\mu}_{\gamma,\rho,1}\cdot\vec{J}\\
\mathcal{T}_4(\gamma,\rho,J,\mu)&=&\int \textrm{d}^{3} \vec{x} f(r) j_{J}(\kappa x) 
\vec{\mathcal{Y}}^{\mu}_{\gamma,\rho,1}\cdot\vec{J},
\end{eqnarray*}
where $Y$ are the spherical harmonics and $\vec{\mathcal{Y}}$ are the
vector harmonics.
The first four operators are those involving the large 
component and are identical with the ones given in \cite{Wal}. Their 
tensor character is $(J,M)$. The last four are  new operators,
i.e. they were ignored in \cite{Wal}, involving  the small component
of the muon wavefunction. Their tensor character is 
$(\gamma,\mu)$. The other indices of the new operators 
identify terms which originate in the multipole expansion to produce 
the correct spherical Bessel function in the integrals. 
The constants appearing in the Hamiltonian 
above are given by the rather lengthy expressions
\begin{eqnarray*}
\alpha(\gamma,\rho,m)&=&\sqrt{\tfrac{3/2+m}{3}} \sqrt{\tfrac{3}{4\pi}}\frac{[\rho]}{[\gamma]}
\left\langle \rho\,1\,\gamma\vert000\right\rangle
\left\langle \rho\,1\,\gamma\vert 0\,m+\tfrac{1}{2}\,m+\tfrac{1}{2}\right\rangle\\
\beta_+(\gamma,\rho,J,m)&=&\sqrt{\tfrac{3/2+m}{3}}
\sqrt{\tfrac{3}{4\pi}}\sqrt{J+1}[J+1]\left\langle 1\,J+1\,\rho\vert000\right\rangle\\
&\phantom{=}&\times \left\langle 1\,J\,\gamma\vert m+\tfrac{1}{2}\,0\,m+\tfrac{1}{2}\right\rangle
W\left(1\,J+1\,\gamma\,1;\rho\,J\right)\\
\beta_-(\gamma,\rho,J,m)&=&\sqrt{\tfrac{3/2+m}{3}}
\sqrt{\tfrac{3}{4\pi}}\sqrt{J}[J-1]\left\langle 1\,J-1\,\rho\vert000\right\rangle\\
&\phantom{=}&\times \left\langle 1\,J\,\gamma\vert m+\tfrac{1}{2}\,0\,m+\tfrac{1}{2}\right\rangle
W\left(1\,J-1\,\gamma\,1;\rho\,J\right)\\
\delta(\gamma,\rho,J,m)&=&\sqrt{\tfrac{3/2-m}{3}} \sqrt{\tfrac{3}{4\pi}}[J]^2\left\langle 
1\,J\,\rho\vert000\right\rangle\\
&\phantom{=}&\times \left\langle 1\,J\,\gamma\vert m-\tfrac{1}{2}\,1\,m+\tfrac{1}{2}\right\rangle
W\left(1\,J\,\gamma\,1;\rho\,J\right)\\
\eta_+(\gamma,\rho,J,m)&=&\sqrt{\tfrac{3/2-m}{3}} \sqrt{\tfrac{3}{4\pi}}\sqrt{J}[J+1]\left\langle 
1\,J+1\,\rho\vert000\right\rangle\\
&\phantom{=}&\times \left\langle 1\,J\,\gamma\vert m-\tfrac{1}{2}\,1\,m+\tfrac{1}{2}\right\rangle
W\left(1\,J+1\,\gamma\,1;\rho\,J\right)\\
\eta_-(\gamma,\rho,J,m)&=&\sqrt{\tfrac{3/2-m}{3}} \sqrt{\tfrac{3}{4\pi}}\sqrt{J+1}[J-1]\left\langle 
1\,J-1\,\rho\vert000\right\rangle\\
&\phantom{=}&\times \left\langle 1\,J\,\gamma\vert m-\tfrac{1}{2}\,1\,m+\tfrac{1}{2}\right\rangle
W\left(1\,J-1\,\gamma\,1;\rho\,J\right)
\end{eqnarray*}
Here we have repeatedly used the standard notation $[J]=\sqrt{2J+1}$. All conventions for the Clebsch-Gordan 
coefficients and W-symbols are those of \cite{Edmonds}.

In the derivation of the Hamiltonian given above we 
have used various selection rules for the Clebsch-Gordan coefficients and 
the W-symbols. Note that some of the 
terms vanish at low angular momenta, since $\gamma$ and $\rho$ must be positive for any 
term to contribute.
The quantity $m$ in the above expressions corresponds to the spin projection of the muon. 
Since we consider unpolarized 
muons we must average over the two values $m=\pm 1/2$. 

For completeness we list here the relevant nuclear currents for the muon capture process. These are
\begin{eqnarray*}
\hat{\rho}_V (\vec{x}) &=& G_E \sum_{j=1}^{A} \tau_+(j)\delta^{(3)}(\vec{x}-\vec{x}_j)\\
\hat{\vec{J}}_V (\vec{x}) &=& \frac{G_E}{2Mi} \sum_{j=1}^{A} \tau_+(j)\left[ 
\delta^{(3)}(\vec{x}-\vec{x}_j)\overrightarrow{\nabla}_{j} - 
\overleftarrow{\nabla}_{j}\delta^{(3)}(\vec{x}-\vec{x}_j)
\right] \\
\phantom{\hat{\vec{J}}_V} &+& \frac{G_M}{2M} \overrightarrow{\nabla} \wedge
\sum_{j=1}^{A} \tau_+(j)\, \vec{\sigma}(j) \delta^{(3)}(\vec{x}-\vec{x}_j) \\
\hat{\rho}_A &=& \frac{G_A}{2Mi} \sum_{j=1}^{A} \tau_+(j)\, \vec{\sigma}\cdot
\left[ 
\delta^{(3)}(\vec{x}-\vec{x}_j)\overrightarrow{\nabla}_{j} - 
\overleftarrow{\nabla}_{j}\delta^{(3)}(\vec{x}-\vec{x}_j)
\right] \\
\phantom{\hat{\rho}_A} &+& \frac{m_\mu G_P}{2M} \overrightarrow{\nabla} \cdot 
\sum_{j=1}^{A} \tau_+(j)\,\vec{\sigma}(j) \delta^{(3)}(\vec{x}-\vec{x}_j)\\
\hat{\vec{J}}_A (\vec{x}) &=& G_A \sum_{j=1}^{A} \tau_+(j)\,\vec{\sigma}(j)\delta^{(3)}(\vec{x}-\vec{x}_j)
\end{eqnarray*}
here $M$ is the nucleon mass, $G_E$ and $G_M$ are the Sachs 
nucleon form factors and $G_A$ is the axial form factor. 
We note that that the usual Fermi and Gamow-Teller transition operators are recovered in the $q\to0$ limit as 
the following multipole components (see, e.g. \cite{Walecka2})
\begin{eqnarray*}
\mathcal{M}_{0,0} &=& \frac{1}{\sqrt{4\pi}} G_V \sum_{j=1}^{A} \tau_+(j) \\
\mathcal{L}_{1,M}&=& \frac{1}{\sqrt{2}}\mathcal{T}_{1,M}^{el}=\frac{i}{\sqrt{12\pi}}
G_A \sum_{j=1}^{A} \tau_+(j)\,\sigma_{1,M}(j)
\end{eqnarray*}
were $G_V$ is the $q\to 0$ limit of the vector coupling form factor, which is often denoted by $F_1(0)$ in 
the literature.

To get the final expression for the total rate one must now evaluate 
the absolute squared matrix element of the Hamiltonian
in the initial and final nuclear states and multiply it by
the two-body phase space factor given in \cite{Wal}.
It is advantegeous  
to group  together the components with like tensor order to better control the interference 
of operators arising from the large and small components.

\end{document}